\documentclass[preprint2]{aastex}
\newcommand{\be}{\begin{equation}}
\newcommand{\ba}{\begin{eqnarray}}
\newcommand{\ee}{\end{equation}}
\newcommand{\ea}{\end{eqnarray}}
\newcommand{\sol}{M_{\sun}}
\newcommand{\ptp}{{\it Prog. Theor. Phys.}}
\newcommand{\ptps}{{\it Prog. Theor. Phys. Suppl.}}
\newcommand{\etal}{{\it et al. }}

\lefthead{Nakamura  }
\righthead{}

\begin{document}

\title{Possible Formation Scenario of the Quark Star of Maximum Mass around
0.7 $\sol$}

\author{Takashi Nakamura }
\affil{Department of  Physics, Kyoto University,
Kyoto 606-8502}

\received{}
\accepted{}

\begin{abstract}
 If there exists the quark star  of  maximum mass $\sim 0.7\sol$
 as suggested by recent Chandra observations, we show 
 that the  general relativistic collapse
of a neutron star of mass  greater than the
maximum mass of the neutron star with angular momentum parameter 
$q\equiv cJ/GM^2 > 1$  may lead to such a strange star.
Here $J$ and $M$ are the angular momentum and the gravitational
mass of the neutron star, respectively. 
Under the cosmic censorship hypothesis, such a star can not
be a black hole directly. The jet formed in the soft core might
explode the outer envelope and leave the quark star of mass  $\sim 0.7\sol$.
The remnant quark star has $\lesssim  10^{53}$ erg rotational energy
so that the formation of the quark star may be related to the central
 engine of GRBs. The detailed numerical simulations are urgent 
to confirm or refute this scenario.
(  version 3 May  2002)
\end{abstract}


\section{Introduction}
 Recently \citet{drake02} have made Deep Chandra LETG+HRC-S observations
of the isolated neutron star candidate RXJ185635-3754 and found
that the X-ray spectrum is well-represented by a $\sim$ 60eV black body.
Their data contain no evidence for pulsation. The derived column density
favors the distance to RXJ185635-3754 to be $D \sim$ 140pc instead of
$D\sim$ 60pc \citep{walter01}. Although \citet{walter02} claim that the
distance is $D \sim 120$pc and the heavy-element atmosphere of \citet{pons02}
should be taken, we here follow the arguments by \citet{drake02}.

{}From the observed flux, the distance and the observed 
temperature of RXJ185635-3754,  only the  radiation
radius defined by $R_\infty \equiv R/\sqrt{1-R_g/R}$ is determined 
since the observed luminosity and the observed temperature 
suffer from the gravitational redshift.
For the given $R_\infty$, $M$ is given as a function of  $R$ as
\be
 R_g = \frac{2GM}{c^2} =R (1-(\frac{R}{R_\infty})^2) .
\ee
 $M$ is positive only for $0 < R < R_\infty$
 and has the maximum value $R_\infty c^2/G3\sqrt{3}=0.7\sol(R_\infty/5.6km)$
at $R=R_\infty/\sqrt{3}$.
 If $R_\infty$ is $\sim 5.6$km$D_{140}$   as suggested by
\citet{drake02}, the maximum mass is $\sim 0.7\sol D_{140}$
where  $D_{140}$ is $D$ in the unit of 140pc.  Therefore $R_\infty$ can
not determine the radius of the star but gives us the upper limit of
the mass as well as the upper limit of the radius of RXJ185635-3754.

X-ray luminosity of RXJ185635-3754 is $\sim 6\times 10^{31}$erg/s$D_{140}^2$.
If this is due to the  accretion from a uniform medium \citep{bondi}
 we have the relation as
\be
   \pi (\frac{2GM}{V^2})^2 \rho V \epsilon = 6\times 10^{31}
{\rm erg s}^{-1}D_{140}^2,
\ee
where $V$ is the velocity of RXJ185635-3754 and 
$\epsilon$ 
is the conversion efficiency of the accreting matter to 
the radiation and is  $ \sim 0.3c^2$ for the quark star with 
the maximum mass $\sim
0.7\sol$\citep{witten}.  From the above equation we have
\be
(\frac{M}{\sol})^2 (\frac{V}{200{\rm kms^{-1}}})^{-3}
(\frac{n}{10^{4}{\rm cm^{-3}}})
(\frac{\epsilon}{0.3c^2})=0.5D_{140}^2
\ee
Since  the proper motion of RXJ185635-3754 is 0.3 arc-second/y
$\sim 200D_{140}{\rm kms^{-1}}$ \citep{walter01} 
\footnote{ Here we assume that the optical counterpart is the same 
object as  RXJ185635-3754 although the luminosity in the optical
band can not be explained by the Rayleigh-Jeans tail of $\sim 60$eV
black body spectrum of RXJ185635-3754 \citep{pons02,drake02,walter02}.} 
the required number density of the interstellar matter is rather high 
 even for $M\sim \sol$ .
While the column density to RXJ185635-3754 is estimated as
$\sim 10^{20}cm^{-2}$  \citep{drake02} so that  the  high density 
region should be local
with the maximum size of $\sim 100$AU. Such a local high density region
of the size $< 100$AU is observed in the interstellar medium using
21cm HI line by \citet{frail94} so that  RXJ185635-3754
may be passing through the high density region now.

If  $M$ is $ \sim 0.1\sol$,  the required density 
of the interstellar matter is so high that
the column density may be in conflict with the observed one
 or we need very small( $\sim$AU) high density region in the
interstellar matter. In the latter case
we expect the change of the luminosity in a time scale of 10 days or so, 
while  only $\sim$2\% change of the luminosity has been observed 
in the 19 months \citep{drake02} so that the mass is much greater 
than  $ \sim 0.1\sol$.
  In this Letter,  assuming the mass and the radius of RXJ185635-3754
are $\sim 0.7 \sol$ and $\sim 3.2$km, we propose a possible formation
scenario of such an object.

\section{Spherically Symmetric Model}

First such a small radius object can not be the neutron star
\citep{lattimer00}.
Even for the softest equation of state of the high density nuclear matter, 
$R_\infty$ is larger than 10km \citep{lattimer00}.
Except for the free neutron case \citep{OV}, the maximum mass of the
neutron star is greater than $ 0.7 \sol$ \citep{lattimer00}, which is
in conflict with the maximum mass of  RXJ185635-3754 from  the
observed $R_\infty$. The strange quark star is one of 
the candidate for such an object like
RXJ185635-3754. In the quark star the baryon number is distributed
among three quarks $u, d$ and $s$ so that  the degenerate energy 
decreases by a factor of 0.89 compared with the proton and  neutron
nuclear matter \citep{witten}. In this sense the strange quark matter
is stable although it contains unstable strange quarks. 
This reason  is essentially
the same as that the unstable neutron exists stably in the stable nucleus.
   In MIT bag model with the large bag constant $B=(245{\rm MeV})^4$
suggested from the finite-temperature QCD lattice data \citep{pesh01} , 
the maximum mass of the quark star $M_{\rm MAX}$ is given by
\ba
M_{\rm MAX}&=&0.7\sol(\frac{B}{(245{\rm MeV})^4})^{-1/2},\\
 R&=&3.9km(\frac{B}{(245{\rm MeV})^4})^{-1/2} , \\
\rho_c&=&1.5\times 10^{16}{\rm g cm^{-3}}(\frac{B}{(245{\rm MeV})^4}),
\ea 
where $R$ and $\rho_c$ are the radius and the central density of the
maximum mass quark star  \citep{witten}.  Note here that the central 
density of the
quark star with $M_{\rm MAX}\sim 0.7\sol$ is much higher than the central
density of the neutron star $\sim10^{15}{\rm g cm^{-3}} $.  Therefore
to make a quark star from the neutron star we must consider
how to increase the central density of the neutron star.
If the bag constant is smaller, for example $B=(145{\rm MeV})^4$ as usual ,
the maximum mass and the radius are $2\sol$ and 11km, respectively.
However, in this case with  $R_\infty=$5.6km , $M$ becomes 
small $\sim 0.15\sol$
which is in conflict with the X-ray luminosity of RXJ185635-3754  as
stated in \S 2 so
that we prefer the larger bag constant. 

A naive way to increase the central density of the neutron star is
to add the mass to the maximum mass neutron star by accretion from
the companion star or the fall back gas in the supernova event \citep{mac}.
However the maximum mass of the neutron star is $1.4\sol \sim 2\sol$
\citep{lattimer00},
which is at least two times larger than the putative maximum mass 
of the quark star relevant to  RXJ185635-3754. The most plausible outcome
of such a collapse is  not the formation of the quark star but the black hole.
In a sense the soft equation of state of the strange quark matter
 ($P=(\rho-4B)/3$)
is favorable to the formation of the strong shock 
in the supernova explosion as shown
by \citet{takahara}. If the equation of state is soft, 
the large density gradient
is accumulated to produce the strong shock while 
for the hard equation of state,
 the small density gradient is enough to produce 
the pressure gradient to support
the falling gas in the collapse.  However the shock is 
the strongest just before the
 formation of the black hole, that is, when the mass of the collapsing star is 
just a bit smaller than the maximum mass of the neutron star
  in the spherically symmetric model \citep{riper}. Since in our case
of the formation of the quark star,  the mass of
 the collapsing matter is much larger than
the maximum mass of the quark star so that we expect the formation of
the black hole.  The detailed numerical simulations might be 
needed to confirm  that the shock can not explode 
the outer envelope  and  the black hole is formed.
\section{Rotating Model}

 I propose another promising scenario.
 First let us review the simulations of general
relativistic collapse of rotating stars.   \citet{nakamura81}
first performed the numerical simulations of general relativistic
collapse of axially symmetric stars of mass $10\sol$ 
leading to the formation of rotating black holes. 
In these simulations the equation of state is
hard with $\gamma =2 $ in the limit of $\rho \gg \rho_{nuc}$
and the rotation is differential with various values  of initial
angular momentum \citep{nakamura81}.
\citet{ns81a} performed similar simulations with
soft equation of state of   $\gamma =4/3 $ for two kinds of rotation laws
with various values  of angular momentum.  The results of these simulations
 crucially depend on one parameter $q$ defined  by
\be
q=\frac{cJ}{GM^2},
\ee 
where $J$ and $M$ are the total angular momentum and the gravitational
mass of the initial rotating star, respectively. When $q\lesssim 1$
the black hole is formed  
 irrespective of initial rotation laws, the equation of state and
the initial density distribution \citep{ns81b}. 
If $q \gtrsim 1$ , the black hole is not formed but the expanding ring,
the expanding disk or the jets appear depending
 on  initial rotation laws, equation of state and 
initial density distribution. This conclusion was  confirmed
 by \citet{sp} and \citet{shibata}.

Physical reason for the above result is as follows; 
Let us consider the collapsing star of mass $M$ and the 
angular momentum $J$.  The radius $R_b$ when the centrifugal
force balances the gravitational force is given by
\be
\frac{GM}{R_b^2}\sim \frac{J^2}{M^2R_b^3}.
\ee
Using $q$, we can express $R_b$ as 
\be
R_b \sim \frac{GM}{c^2}q^2.
\ee
Equation (9) means that if $q \lesssim 1$,
$R_b$ is smaller than the black hole, which suggests
that when the rotation becomes important,  the star itself
is already inside the black hole. Therefore it is natural
that the black hole is formed  irrespective of initial rotation laws, 
the equation of state and the initial density distribution.

In the sense of general relativity, the results of the simulations
support the cosmic censorship hypothesis \citep{penrose} which says that 
 the singularity does not exist outside the event horizon in nature.
Under this hypothesis,   
the Kerr metric with $a/m=q <1$  is known to be  unique as the
ultimate  space-time of the gravitational collapse \citep{robinson}.  
 Note here that $q$ defined in Eq. (7)
and $a/m$ in Kerr metric are the same quantity.
 If the whole of the collapsing star with  $q \gtrsim 1$ becomes a black hole, 
that is the counter example of the cosmic censorship hypothesis. 
However numerical simulations have not yet found  such a case.

In relation to the formation of the quark star, \citet{ns81a,ns81b} showed an
interesting case of the jet formation. In their model of A146
which is the general relativistic collapse of the rotating star
with $\gamma =4/3$ equation of state and the initial $q=1.46$,
the jet appeared in the final phase of the collapse. This is in accord
with the result of \citet{riper,takahara}. In the rotating collapse, 
when the size of the star  becomes $\sim R_b$ in Eq. (9), the collapse
perpendicular to the rotation axis halts while the collapse along
the rotation axis proceeds. When the equation of state is soft, the
large density gradient accumulates to produce the jet. While if
the equation of state is hard, this is not so. In reality \citet{nakamura87}    
performed the simulations of the accreting neutron star with rotation. They
 used the equation of state with  the maximum mass $1.4\sol$ and
simulated the collapse of the rotating neutron star of mass $1.9\sol$.
 They found no jets.
The final result was the differentially rotating neutron star.

Recently \citet{baumgarte} computed the equilibria of 
differentially rotating neutron stars. They stated that some of
the models exceed the Kerr limit, that is , $q > 1$.  If we add
certain amount of matter to exceed the maximum mass neutron star keeping
$q > 1$, the collapse will start. At $R_b$,  the density is estimated as
\be
\rho_b \sim \frac{M}{fR_b^3}=6.7\times 10^{16}{\rm g cm^{-3}}f^{-1}q^{-6}
(\frac{M}{3\sol})^{-2},
\ee
where $f\sim 1$ is the form factor of the rotating star.
Since $\rho_b$ can be in the quark phase for appropriate parameters,
the equation of state in the central core can be  soft
 at the centrifugal bounce. Then the matter will collapse along 
the rotational axis to form the jet like A146 in  \citet{ns81a,ns81b}.
If the jet is strong enough to explode the outer envelope
and leave the central core of mass $\sim 0.7\sol$,  the strange
 quark star might be formed. The important point here is that
if the cosmic censorship hypothesis is correct as suggested by numerical
simulations,  the whole of the rotating neutron star  of mass  greater than the
maximum mass of the neutron star with $q >1 $ can 
not be a black hole directly while if  $q =0$ 
it is possible to be a black hole directly.

The detailed numerical simulations are urgent
to confirm or refute the present scenario  of formation of 
 the strange quark star of maximum mass around 0.7 $\sol$.

\section{Discussion}
 \citet{olinto} showed that a seed of strange matter in a neutron star
will convert the star into a strange star. As a seed,  the quark nugget formed
at the QCD phase transition in the early universe \citep{witten} 
can be considered.  
Then the capture 
cross section of the  quark nugget by the neutron star is given by
\be
\sigma \sim \pi R_{\rm NS}(\frac{GM_{\rm NS}}{V_{\rm rel}^2})
\sim 10^{18}{\rm cm^2}(\frac{V_{\rm rel}}{100{\rm kms^{-1}}})^{-2},
\ee 
where $R_{\rm NS}$, $M_{\rm NS}$ and $V_{\rm rel}$ are 
the radius of the neutrons star,
 the mass of the neutron star and the relative velocity, respectively. 
In order that at least a single quark nugget should be captured in the
age of the universe ($ t_U$) for a certain neutron star , 
the number density of the quark nugget ($n_{\rm nugget}$)
should be
\be
n_{\rm nugget}  > \frac{1}{\sigma V_{\rm rel} t_U}=10^{-42}{\rm cm^{-3}}
 (\frac{V_{\rm rel}}{100{\rm kms^{-1}}})(\frac{t_U}{10^{10}y})^{-1}.\nonumber
\ee
Since the mass density of the quark nugget is at most
the same as the dark matter density, each mass of the
quark nugget $m_{\rm nugget}$ should be smaller than $\sim 10^{12}$g.
However such a small mass quark nugget may not be formed due to
the evaporation ($m_{\rm nugget} > 10^{20}$g for survival; \citet{bhat}).  
Therefore to explain the possible strange quark star of RXJ185635-3754,
 the quark nugget should have been captured 
by chance by the progenitor neutron star.

The putative mass of $\sim 0.7\sol$ is just the mass of the typical
white dwarf \citep{brag}. Therefore if we can compress the
white dwarf of density $\sim 10^7{\rm g cm^{-3}}$ to 
$\sim 10^{16}{\rm g cm^{-3}}$
, the strange quark star of mass $0.7\sol$  might be formed.
A possible process is the tidal pinching proposed by \citet{luminet}.
If the moderate massive black hole of mass $\sim 2000\sol$
exists, a white dwarf passing inside the tidal radius of the
black hole may be pinched and the density increases 
by more than a factor 50 \citep{luminet}. Although  the existence of 
such an intermediate mass
black hole has been suggested \citep{matsumoto},  
the number density of the white dwarfs is small 
$\sim 0.01{\rm pc}^{-3}$ so that the pinching number 
in the age of the universe
 for a single intermediate mass
black hole
is very small as
\ba
&&N_{pinch}\sim n_{WD}(\frac{6GM_{BH}}{c^2}) (\frac{GM_{BH}}
{V_{\rm rel}^2})V_{\rm rel}t_U \nonumber\\
&&\sim10^{-8}(\frac{n_{WD}}{0.01{\rm pc}^{^3}})(\frac{M_{BH}}{2000\sol})^2
(\frac{V_{\rm rel}}{100{\rm kms^{-1}}})^{-1}(\frac{t_U}{10^{10}y}).\nonumber
\ea
Therefore to explain the possible strange quark star of RXJ185635-3754,
 the white dwarf should have been tidally pinched  by the intermediate mass
black hole by chance.

In any scenario, if the quark star of mass $M_{qs}\sim 0.7\sol$ is formed,
the huge binding energy $\sim 4\times 10^{53}$erg is liberated.
The major  part of this energy may be used to explode the outer envelope
of the progenitor neutron star. If , however, only $\sim 10\%$ of this energy
is deposited as the kinetic energy, it is enough to explain the
luminous supernova such as SN1998bw. In the rotating model, the remnant
quark star should have an initial value of $q_{qs} \lesssim 1$.
Then the rotational energy is estimated as
$E_{rot}\sim (GM_{qs}/Rc^2)^2q_{qs}^2M_{qs}c^2\sim 10^{53}$erg.
This energy is larger than or comparable to the energy needed for
GRB  so that the formation of the quark star may be 
related to the central engine of GRB.

\acknowledgments
I would like to thank Chiba, Inutsuka, Tatsumi and Tsuru for
useful comments.  This work was supported in part by
Grant-in-Aid of Scientific Research of the Ministry of Education,
Culture, and Sports, No.14047212 and No.14204024.

\end{document}